\documentclass[twocolumn]{aastex631}

\usepackage{color, soul}
\usepackage{etoolbox}
\makeatletter % Workaround to only color the year for cite commands
 \patchcmd{\NAT@citex}
   {\@citea\NAT@hyper@{%
     \NAT@nmfmt{\NAT@nm}%
     \hyper@natlinkbreak{\NAT@aysep\NAT@spacechar}{\@citeb\@extra@b@citeb}%
     \NAT@date}}
   {\@citea\NAT@nmfmt{\NAT@nm}%
   \NAT@aysep\NAT@spacechar\NAT@hyper@{\NAT@date}}{}{}

 \patchcmd{\NAT@citex}
   {\@citea\NAT@hyper@{%
     \NAT@nmfmt{\NAT@nm}%
     \hyper@natlinkbreak{\NAT@spacechar\NAT@@open\if*#1*\else#1\NAT@spacechar\fi}%
       {\@citeb\@extra@b@citeb}%
     \NAT@date}}
   {\@citea\NAT@nmfmt{\NAT@nm}%
   \NAT@spacechar\NAT@@open\if*#1*\else#1\NAT@spacechar\fi\NAT@hyper@{\NAT@date}}
   {}{}
\makeatother

\newcommand{\Athena}{\textsc{Athena++}}
\newcommand{\TE}{\langle T\rangle_E}
\newcommand{\TV}{\langle T\rangle_V}
\newcommand{\rhog}{\rho g}
\newcommand{\rhov}{\rho v}
\newcommand{\EV}{\langle E_{\rm int}\rangle_V}
\newcommand{\lP}{\langle P}
\newcommand{\LeqT}{$L/4\pi r^2=\sigma T_{\mathrm{eff},F}^4$}

\newtoggle{referee}            %
\togglefalse{referee}          %

\iftoggle{referee}{%           %
  \usepackage{color,soul}    %
  \soulregister\citet7
  \soulregister\citep7
  \soulregister\ref7
  \newcommand{\hlm}[1]{%
      \hl{#1}%  %
  }                          %
}{%                            %
  \newcommand{\hlm}[1]{{#1}} %
}                              %
\usepackage{xcolor}
\usepackage{amsmath}
\usepackage{CJK}

\received{May 11, 2023}
\revised{May 26, 2023}
\accepted{????}
\submitjournal{ApJL}

\shorttitle{Turbulence Supported Envelopes}
\shortauthors{Schultz, Bildsten, and Jiang}

\graphicspath{{./}{figures/}}
\begin{document}
\begin{CJK*}{UTF8}{gbsn}

\title{Turbulence Supported Massive Star Envelopes 
 \footnote{Released on ??, ??, 2023}}
%%%%%%%%%%%%%%%%%%%%%%%%%%%%%%%%%%%%%%%%%%
\correspondingauthor{W. C. Schultz}
\email{wcschultz@physics.ucsb.edu}

\author[0000-0003-1796-9849]{William C. Schultz}
\affiliation{Department of Physics, University of California, Santa Barbara, CA 93106, USA}

\author{Lars Bildsten}
\affiliation{Department of Physics, University of California, Santa Barbara, CA 93106, USA}
\affiliation{Kavli Institute for Theoretical Physics, University of California, Santa Barbara, CA 93106, USA}

\author[0000-0002-2624-3399]{Yan-Fei Jiang(姜燕飞)}
\affiliation{Center for Computational Astrophysics, Flatiron Institute, New York, NY 10010, USA}

\begin{abstract}
The outer envelopes of massive ($M\gtrsim10\,M_{\odot}$) stars exhibit large increases in opacities from forests of lines and ionization transitions (particularly from iron and helium) that trigger near-surface convection zones.
One-dimensional models predict density inversions and supersonic motions that must be resolved with computationally intensive 3D radiation hydrodynamic (RHD) modeling. 
Only in the last decade have computational tools advanced to the point where ab initio 3D models of these turbulent  envelopes can be calculated, enabling us to present five 3D RHD \Athena\ models (four previously published and one new 13$M_{\odot}$ model).
When convective motions are sub-sonic, we find excellent agreement between 3D and 1D velocity magnitudes, stellar structure, and photospheric quantities.
However when convective velocities approach the sound speed, hydrostatic balance fails as the turbulent pressure can account for 80$\%$ of the force balance.
As predicted by Henyey, we show that this additional pressure support leads to a modified temperature gradient which reduces the superadiabaticity where convection is occurring.
In addition, all five models display significant overshooting from the convection in the Fe convection zone.
As a result, the turbulent velocities at the surface are indicative of those in the Fe zone.
There are no confined convection zones as seen in 1D models. 
In particular, helium convection zones seen in 1D models are significantly modified.
Stochastic low frequency brightness variability is also present in the 13$M_{\odot}$ model with comparable amplitude and characteristic frequency to observed stars.
\end{abstract}

\keywords{Stellar physics (1621), Stellar convective zones (301), Stellar surfaces (1632), Stellar structures (1631)}

%%%%%%%%%%%%%%%%%%%%%%%%%%%%%%%%%%%%%%%%%%
\section{Introduction} \label{sec:intro}
The opacity ($\kappa$) increases associated with Fe and He excite near-surface convection zones (NSCZs) in massive star envelopes \citep{Cantiello2009} across the Hertzsprung Russell diagram (HRD).
In many cases, 1D modeling of these NSCZs predict trans-sonic convective velocities ($v_{\mathrm{c}}$), and as the luminosity ($L$) approaches the Eddington value, $L_{\mathrm{Edd}}=4\pi GMc/\kappa$, produce unstable density inversions \citep{Joss1973}.
These properties present a serious challenge to modeling the nature of convection in NSCZs with one-dimensional (1D) approaches such as mixing length theory (MLT) \citep{BohmVitense1958}, and motivate our extended suite of 3D radiation hydrodynamic (RHD) simulations using \Athena\ \citep{Stone2020,Jiang2021}.

The 3D NSCZs show significant fluctuations in density ($\rho$) and radiative flux ($F_{\mathrm{r}}$) that reduce the radiation pressure support as a result of porosity \citep{Schultz2020}.
These density fluctuations also combine with trans-sonic velocities and propagate through the stellar photosphere to produce an optically thick wind in sufficiently high $L$ models \citep{Jiang2015,Jiang2018}.
As the plumes become optically thin near the photosphere, they impart significant Doppler broadening on the escaping photons, naturally generating the long-observed microturbulence measured in spectral line widths of massive stars \citep{Cantiello2009,Schultz2023}.
The 3D surface convection also generates stellar brightness variability with amplitudes and frequencies \citep{Schultz2022} similar to observed massive stars \citep{Bowman2020a} in recent photometric surveys \citep[e.g. TESS,][]{Ricker2015} confirming 1D analyses \citep{Cantiello2021}.

In this paper, we utilize the variety in the optical depth at the Fe opacity peak ($\tau_{\mathrm{Fe}}$) in our \Athena\ simulation suite to elucidate the diversity of 3D convection from regimes where energy is transported via convection to the more ``lossy" realm where radiative energy transport dominates even though convection is vigorous. 
As discussed in \cite{Goldberg2021, Jermyn2022RN}, radiative transport becomes important when $\tau_{\mathrm{Fe}}$ is less than the optical depth,
\begin{equation} \label{eq:tau_crit}
    \tau_{\rm crit}=\frac{P_{\mathrm{r}}c}{(P_{\mathrm{r}}+P_{\mathrm{g}})v_{\mathrm{c}}},
\end{equation}
where $P_{\mathrm{r}}$ is the radiation pressure, $P_{\mathrm{g}}$ is the gas pressure, and $c$ is the speed of light.
When $\tau_{\mathrm{Fe}}\ll\tau_{\mathrm{crit}}$, convective plumes lose heat from radiative diffusion as they travel upwards causing a reduction in convective efficiency, and a smaller convective flux. 
For massive main sequence and Hertzsprung Gap stars, $\tau_{\mathrm{crit}}\gtrsim1000$, allowing radiatively leaky NSCZs to exist inside an optically thick envelope.
Such a realm of convection is implicitly included in the early works of \cite{Henyey1965,Ludwig1999,Kuhfuss1986}, and those early estimates of the impact of the radiative losses guided our exploration.

Our focus here is two fold.
First, we exhibit that for models where convective transport dominates, the trans-sonic velocities apply a significant pressure, $P_{\mathrm{turb}}\propto\rhov_{\mathrm{c}}^2$, that affects the outer envelope structure much like that seen in 3D models of red supergiants \citep[e.g.][]{Goldberg2022}. 
The temperature profiles in these 3D models can then be well explained (and predicted) by incorporating the impact of turbulent pressure using the \cite{Henyey1965} model.
\hlm{Similar work has shown $P_{\mathrm{turb}}$ plays a crucial role in the explosions of core-collapse supernovae} \citep[e.g.][]{Couch2015}.
Secondly, for hotter massive stars on the main sequence, the effects of turbulent pressure are much less \citep[e.g.][]{Grassitelli2015}.
However, these 3D RHD models exhibit convective motions far outside of the regions conventionally defined by 1D models and have detectable photometric variability.

%%%%%%%%%%%%%%%%%%%%%%%%%%%%%%%%%%%%%%%%%%
\section{Hydrostatic Balance in 3D Models of Turbulent Envelopes}
\begin{figure}[t]
%\vspace*{-2.0 cm}
\begin{center}
\includegraphics[width=0.45\textwidth]{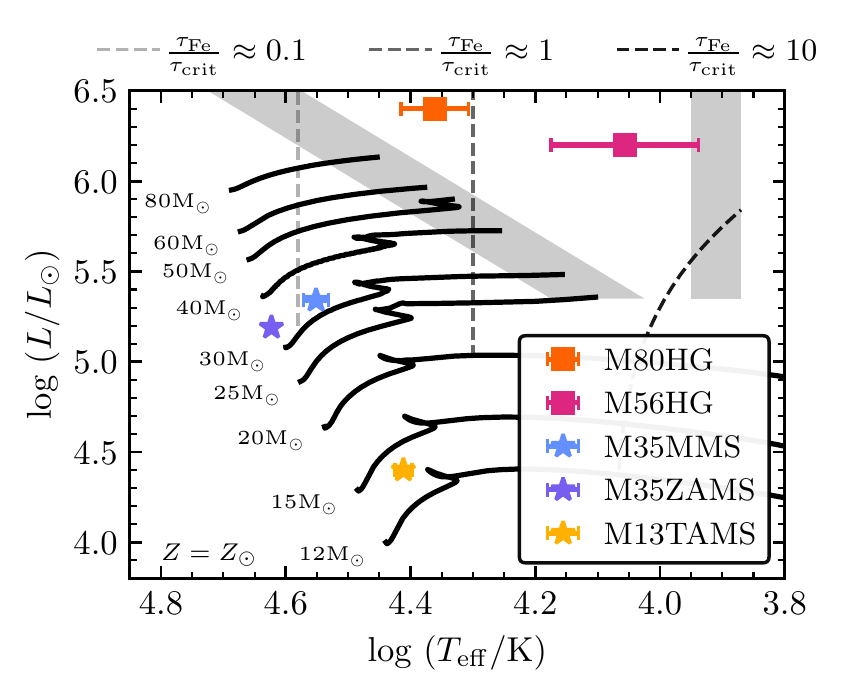} 
\vspace*{-.4 cm}
\caption{HRD showing the five 3D models (colored points).
Squares represent models from \cite{Jiang2018} while the stars denote models from \cite{Schultz2022} and this work.
The errorbars show the difference between definitions of $T_{\mathrm{eff}}$ described in \S~\ref{sec:1d_models}.
The black lines are MESA models from \cite{Cantiello2021}.
The vertical dashed lines are approximate contours of $\tau_{\mathrm{Fe}}/\tau_{\mathrm{crit}}$ as calculated from the MESA models.
Grey shaded regions show the S Dor and LBV outburst instability strips from \cite{Smith2004}.
}
\label{fig:HR}
\end{center}
\end{figure}

Five 3D RHD \Athena\ solar metallicity stellar envelope models were run to steady-state equilibrium, defined by reaching thermal equilibrium beneath the FeCZ.
Table~\ref{tab:3D_models} details relevant quantities with the model names denoted by the core mass and phase of evolution: zero-age, middle of, and terminal age main sequence (ZAMS, MMS, TAMS respectively), and Hertzsprung Gap (HG).
The new model, M13TAMS, consists of a narrow wedge spanning $>20$ scale heights in both angular directions and a radial range that includes both the radiative region of the envelope below the CZs and the regions beyond the photosphere.
Figure~\ref{fig:HR} shows the location in the HRD of these models relative to MESA \citep[Modules for Experiments in Stellar Astrophysics;][]{Paxton2011,Paxton2013,Paxton2015,Paxton2018,Paxton2019,Jermyn2023} tracks.

\begin{splitdeluxetable*}{lcccccccccBlccccccccc}
\tabletypesize{\footnotesize}
\tablenum{1}
\tablecaption{3D Model Properties \label{tab:3D_models}}
\tablewidth{0pt}
\tablehead{
\colhead{Name} & \multicolumn{2}{c}{Masses} & \multicolumn{2}{c}{Temperature} & \multicolumn{2}{c}{Luminosities} & \colhead{Angular Size} & \multicolumn{2}{c}{Optical Depth} & \colhead{Name} & \multicolumn{2}{c}{MESA Velocities} & \multicolumn{3}{c}{3D RMS Velocities}  & \multicolumn{4}{c}{Scale Heights} \\
\nocolhead{Name} & \colhead{$M_{\mathrm{core}}$} & \colhead{$M_{\mathrm{env}}$} & \colhead{$T_{\mathrm{eff},F}$ $^{\mathrm{a}}$} & \colhead{$T_{\mathrm{eff}}$ $^{\mathrm{b}}$} & \colhead{$L$} & \colhead{$\Gamma_{\rm Edd,\,Fe}$ $^{\mathrm{c}}$} & \colhead{$\Omega_{\mathrm{sim}}$} & \colhead{$\tau_{\mathrm{Fe}}$} & \colhead{$\frac{\tau_{\mathrm{Fe}}}{\tau_{\mathrm{crit}}}$} & \nocolhead{Name} & \colhead{$v_{\mathrm{MESA, Fe}}$} & \colhead{$v_{\mathrm{MESA, He}}$} & \colhead{$v_{\mathrm{Fe}}$} & \colhead{$v_{\mathrm{He}}$} & \colhead{$v_{F}$} & \colhead{$H_{\mathrm{Fe}}$} & \colhead{$H_{F}$} & \colhead{$H_{\mathrm{tot, Fe}}$} & \colhead{$H_{\mathrm{tot}, F}$} \\
\nocolhead{Name} & \colhead{(M$_{\odot}$)} & \colhead{(M$_{\odot}$)} & \colhead{($10^3\,$K)} & \colhead{($10^3\,$K)} & \colhead{(log($L$/L$_{\odot}$))} & \nocolhead{Eddrat)} & \colhead{($\mathrm{sr}/\pi$)} & \nocolhead{tau} & \nocolhead{tau} & \nocolhead{Name} & \colhead{(km$\,$s$^{-1}$)} & \colhead{(km$\,$s$^{-1}$)} & \colhead{(km$\,$s$^{-1}$)} & \colhead{(km$\,$s$^{-1}$)} & \colhead{(km$\,$s$^{-1}$)} & \colhead{(R$_{\odot}$)} & \colhead{(R$_{\odot}$)} & \colhead{(R$_{\odot}$)} & \colhead{(R$_{\odot}$)}
}
%\decimalcolnumbers
\startdata
M35ZAMS $^1$ & 35 & 0.004 & 42 & 42 & 5.2 & 0.82 & 0.0013 & 515 & 0.02 & M35ZAMS & 20 & 0 & 6.2 & 8.0 & 8.9 & 0.07 & 0.01 & 0.07 & 0.01\\
M35MMS $^2$& 35 & 0.027 & 34 & 37 & 5.4 & 0.97 & 0.04 & 1,084 & 0.25 & M35MMS & 44 & 0 & 57 & 145 & 148 & 0.25 & 0.09 & 0.28 & 0.13 \\
M13TAMS & 13 & 0.019 & 25 & 27 & 4.4 & 0.55 & 0.0096 & 3,487 & 0.24 & M13TAMS & 11 & 0.01 & 8.3 & 12 & 19  & 0.11 & 0.02 & 0.11 & 0.02\\
M80HG $^3$& 80 & 0.032 & 26 & 20 & 6.4 & 3.23 & 1.41 & 4,300 & 2.3 & M80HG & - & - & 145 & 170 & 166 & 3.99 & 2.58 & 5.74 & 8.36 \\
M56HG $^4$& 56 & 0.036 & 15 & 9 & 6.2 & 3.38 & 1.41 & 25,686 & 8.1 & M56HG & - & - & 81 & 132 & 161 & 13.34 & 4.83 & 16.45 & 31.12 \\
\enddata

\tablenotetext{\mathrm{a}}{Defined as the average temperature at the radius, $r$, that satisfies \LeqT.}
\tablenotetext{\mathrm{b}}{Defined as the average temperature at the location where $\langle\tau\rangle=1$.}
\tablenotetext{\mathrm{c}}{Eddington ratio at the Fe opacity peak.}
\tablenotetext{1}{Model T42L5.0 in \cite{Schultz2022, Schultz2023}.}
\tablenotetext{2}{Model T32L5.2 in \cite{Schultz2022} and T35L5.2 in \cite{Schultz2023}.}
\tablenotetext{3}{Model T19L6.4 in \cite{Jiang2018, Schultz2020,Schultz2023}.}
\tablenotetext{4}{Model T9L6.2 in \cite{Jiang2018,Schultz2020}.}

\end{splitdeluxetable*}

\begin{figure}[ht]
%\vspace*{-2.0 cm}
\begin{center}
 \includegraphics[width=0.45\textwidth]{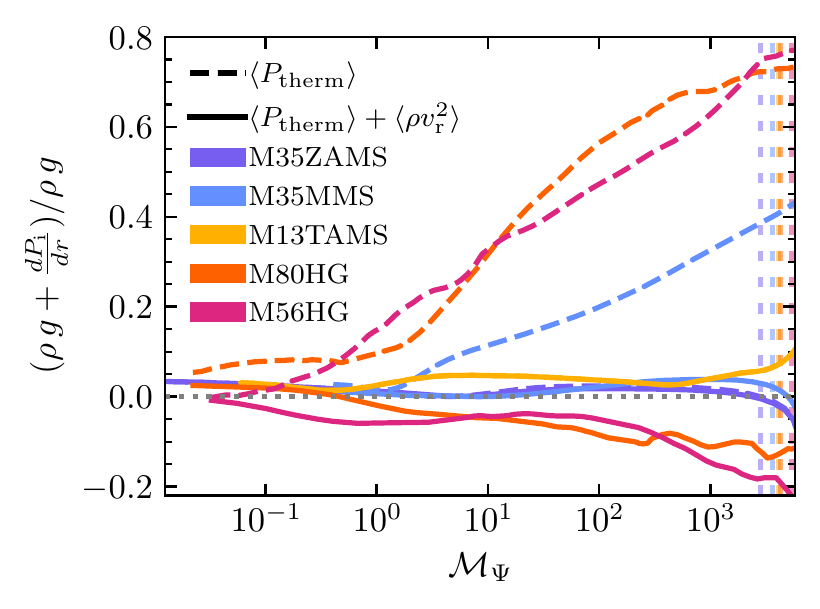} 
\vspace*{-.4 cm}
\caption{Fractional difference from HB for the average pressure of 3D models versus $\mathcal{M}_{\Psi}$ including and excluding turbulent pressure (solid and dashed respectively).
Vertical dashed lines on the right denote the location where the energy flux matches the photospheric criteria of $\sigma T^4$ (e.g. $T=T_{\mathrm{eff},\,F}$ from Table~\ref{tab:3D_models}).
}
\label{fig:mom_check}
\end{center}
\end{figure}

\cite{Schultz2020} showed that these 3D RHD models exhibit correlations that reduce the radiation pressure support, motivating our investigation of hydrostatic balance (HB).
To quantify HB, the relative deviations, $(\rhog+dP/dr)/\rhog$ are plotted in Figure~\ref{fig:mom_check} using different choices for the pressure support.
The x-axis is the pseudo-Mach number defined in \cite{Schultz2020} and increases outwards monotonically.

Models M13TAMS and M35ZAMS are in HB with $P=\lP_{\mathrm{therm}}\rangle_V$, deviating by less than $5\%$ throughout their envelopes.
For the other three models, however, HB fails with M35MMS, M80HG, and M56HG reaching upwards of $41\%$, $73\%$, and $77\%$ discrepancies at their surfaces respectively.
These large deviations cannot be explained by the optically thick winds of M80HG and M56HG, as the advective velocity associated with the wind mass loss is less than the turbulent velocities inside the photosphere.
The positive discrepancy indicates additional pressure support is needed to reach force balance (FB) for these turbulent models.

As the RMS velocities of the turbulence approach the sound speed, turbulent pressure given by $P_{\mathrm{turb}}\equiv\langle\rhov_{\mathrm{r}}^2\rangle_V$ provides support.
The choice of the radial velocity component, $v_{\mathrm{r}}$, is motivated by expanding the radial component of the 3D spherical polar momentum equation where a $\rhov_{\mathrm{r}}^2$ term assists in balancing $\rhog$ \citep{Goldberg2022}.
This additional pressure does not alter the comparison for M35ZAMS and M13TAMS as $P_{\mathrm{turb}}\ll P_{\mathrm{therm}}$ in these models.
However the other three models show significant improvements in FB when $P_{\mathrm{turb}}$ is added.
The deviations of M35MMS reduce to the same $5\%$ as the two less turbulent models.
The models with significant optically thick winds only experience $<20\%$ deviations within their photospheres, a factor of 4 improvement, and moreover, the deviations are in the expected direction.
Accounting for $P_{\mathrm{turb}}$ in the luminous and massive models is required to correctly balance gravity in 1D stellar evolution models. 

%%%%%%%%%%%%%%%%%%%%%%%%%%%%%%%%%%%
\section{Comparing Averages of 3D Models to 1D Profiles}
As 3D RHD models of massive star envelopes are computationally expensive, 1D models remain the most effective way to understand their evolution.
However 1D models require approximations that are not always verified with physically motivated 3D models.
By comparing the 1D averages of the existing 3D models we can verify the effectiveness of current 1D approximations.
Spherical averages of the 3D models were calculated and compared to MESA models chosen to match their core masses and luminosities.
Only the three wedge-like models (M35ZAMS, M35MMS, and M13TAMS) were able to be precisely matched as the global models (M80HG and M65H) have luminosities boosted by $\approx50\%$, to yield faster convergence of their turbulent motions.

\subsection{Extracting 1D Models from 3D Wedges} \label{sec:1d_models}
Distilling the turbulent 3D RHD models into 1D analogs is not as simple as using a volume weighted average.
Using $\langle\rho\rangle_V$ and $\TV$ to calculate the thermal pressure results in factor of 2 differences from $\lP_{\rm therm}\rangle_V$ near the surface, motivating a reconsideration of conserved quantities when averaging 3D models to yield 1D profiles.

We chose the mass and internal energy contained within each radial shell to be the conserved quantities in the 3D to 1D translation.
As the gas and radiation temperatures in each 3D cell deviate by $<2\%$, only one temperature is needed when calculating the internal energy there-in,
\begin{equation} \label{eq:Eint}
    E_{\mathrm{int}}=\frac{3}{2}P_{\mathrm{g}}+3P_{\mathrm{r}}=\frac{3k_{\mathrm{B}}}{2\mu m_{\mathrm{p}}}\rho T+a_{\mathrm{r}}T^4,
\end{equation}
where $\mu$ is the mean molecular weight.
Thus we utilized the volume weighted averages of the internal energy, $\EV$, and density, $\langle\rho\rangle_V$, to calculate a self-consistent average temperature, $\TE$, using Eq.~\ref{eq:Eint}.
Replacing $\TV$ with $\TE$, reduces the fractional difference of the internal energy ($|E_{\mathrm{int}}(\langle\rho\rangle_V,\langle T\rangle_i)-\EV|/\EV$) from upwards of $60\%$ to numerical error. 
Using $\TE$ to estimate the thermal pressure results in an average deviation of $10^{-3}$ with the maximal deviation staying $<5\%$ for all the models.
This is a significant improvement compared to the average and maximal deviations of $5\%$ and $60\%$ respectively when using $\TV$.
Because of these improved agreements, $\TE$ and $\langle\rho\rangle_V$ are taken to be the 1D averages of the 3D models.  

\begin{figure*}[ht]
%\vspace*{-2.0 cm}
\begin{center}
 \includegraphics[width=\textwidth]{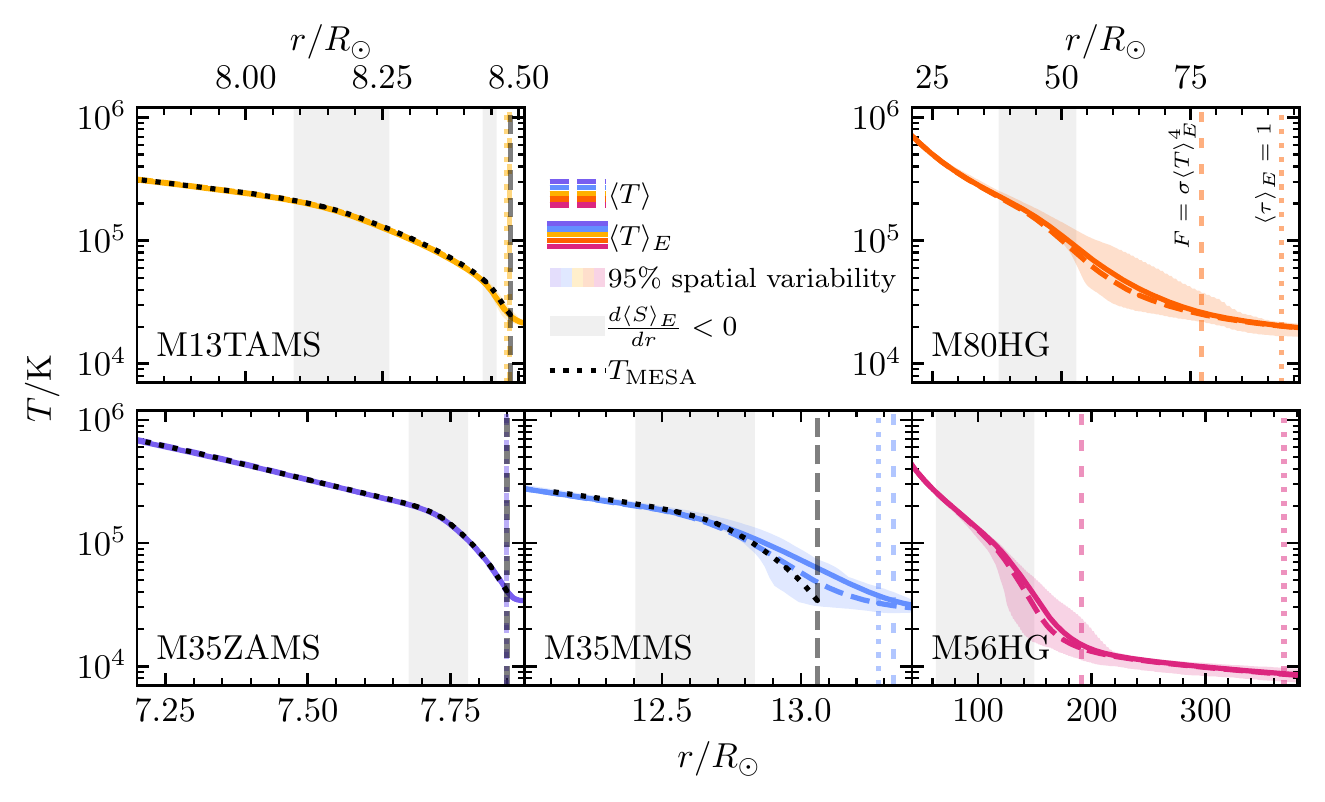} 
\vspace*{-0.75 cm}
\caption{Comparison of radial profiles of the volume weighted average temperature, $\TV$ (dotted lines), and the temperature extracted from the internal energy, $\TE$ (solid lines), for the five models.
The shaded color regions show the $95\%$ spatial variability in a single temporal snapshot.
Vertical colored lines show the radii where $\langle\tau\rangle_E=1$ (dotted) and $L/4\pi r^2=\sigma\TE^4$ (dashed).
The grey shaded regions denote where the average entropy gradient is negative, signifying a CZ.
Temperature profiles from MESA models of similar stars to M13TAMS, M35ZAMS, and M35MMS are plotted in black dotted lines with their photospheric radii shown by the vertical grey dashed lines.
}
\label{fig:new_Ts}
\end{center}
\end{figure*}

Figure~\ref{fig:new_Ts} compares $\TV$ and $\TE$ along with the typical spatial variation of $T$ in a single temporal snapshot.
The plotted regions of the models span from the innermost radius to reach thermal equilibrium to the outer radius where $\langle\tau\rangle_E=1$ ($\langle\tau\rangle_E(r)\equiv\int_r^{\infty}\kappa(\TE,\langle\rho\rangle_V)\langle\rho\rangle_V dr'$).
In M13TAMS and M35ZAMS, $\TE$ deviates from $\TV$ by less than $1\%$ while it changes by upwards of $23\%$, $20\%$, and $13\%$ in M35MMS, M56HG , and M80HG respectively signifying the turbulence generates correlations between $T$ and $\rho$.

The overlapping or distinct locations of photosphere definitions, highlighted in Figure~\ref{fig:new_Ts}, arise from the varied topographies of these 3D RHD models.
The 1D nature of M35ZAMS and M13TAMS leads to the location where $\langle\tau\rangle_E=1$ coinciding within one radial cell of where $F=\sigma\TE^4$.
In contrast, the other three models display convective motions that reach or surpass the sound speed, resulting in strong turbulent motions at the photosphere and complex surface topography.
The trans-sonic turbulence of M35MMS results in a $10\%$ discrepancy in the predicted $T_{\mathrm{eff}}$ based on the choice of photospheric definitions, with the location where $\langle\tau\rangle_E=1$ lying inside the flux defined surface.
When turbulence becomes super-sonic, as in M80HG and M56HG, the different $T_{\mathrm{eff}}$ definitions deviate by $30\%$ and $40\%$ respectively with $F=\sigma\TE^4$ occurring well below the location where $\langle\tau\rangle_E=1$.

This contrast highlights the inherently 3D nature of these models.
The turbulence causes complex surface topography with the radially integrated $\tau=1$ surface spanning up to twice the stellar radius and encompassing both photospheric definitions plotted in Figure~\ref{fig:new_Ts} \citep{Schultz2023}.
The large topographic features yields factor of two temperature fluctuations, containing both definitions of $T_{\mathrm{eff}}$ in Table~\ref{tab:3D_models} and breaking the notion of a single $T_{\mathrm{eff}}$ or single radius photosphere.
In order to show an approximation for the surface of the 1D averages of these 3D RHD models, $F=\sigma\TE^4$ is chosen to define the 1D photosphere and future figures are truncated there.

\subsection{Comparing to MESA Models}
Matching the solar metallicities of the 3D RHD models and adding exponential core overshooting with $f=0.0014$, $f_0=0.004$ to smooth the HRD tracks, pre-main sequence models for each core mass are modeled to ZAMS using the latest release of MESA (r22.11.1).
These ZAMS models are then evolved utilizing the \cite{Henyey1965} MLT option, with the default parameters ($\alpha=2,\,y=1/3,\,\nu=8$) until the $\TE$, $\langle\rho\rangle_V$, and $r$ of the base of the 1D analogs of the 3D RHD models are matched.
All three models matched these conditions to within $5\%$ as the base radii are well below the Fe opacity peak in a radiative region. 
\hlm{The inlists used to create these models are available on Zenodo under an open-source 
Creative Commons Attribution license:} 
\dataset[doi:10.5281/zenodo.7972070]{https://doi.org/10.5281/zenodo.7972070}.

The $T$ (see Figure~\ref{fig:new_Ts}) and $\rho$ MESA profiles agree with M35ZAMS and M13TAMS at all radii, including the photosphere.
This is the first direct confirmation that 3D RHD \Athena\ simulations with convective turbulence agree with 1D MESA models and highlights a region of the HRD where 1D stellar evolution models are sufficient to capture the stellar structure as well as the photospheric radii and temperatures.
This region coincides with $\tau_{\mathrm{Fe}}\ll\tau_{\mathrm{crit}}$ or $L\ll L_{\mathrm{Edd}}$ as in these limits the convection does not carry significant energy and cannot develop trans-sonic turbulence.
In contrast, the MESA analog to M35MMS is slightly more compact and hotter, with $r_{\mathrm{photo}}$ and $T_{\mathrm{eff}}$ $\approx2\%$ different than the 3D estimates.
Additionally, the 3D estimate exhibits a shallower temperature gradient through and above the FeCZ, causing the change in $r_{\mathrm{photo}}$ and $T_{\mathrm{eff}}$.
Despite the lack of MESA models to compare to, these discrepancies are likely to increase in the models that exhibit stronger turbulent motions (i.e. models with more massive cores, higher luminosities, and with $\tau_{\mathrm{Fe}}\gtrsim\tau_{\mathrm{crit}}$).
It is clear that new prescriptions of turbulent heat transport are required to improve the 1D stellar evolutionary models in this regime.

\begin{figure*}[ht]
%\vspace*{-2.0 cm}
\begin{center}
 \includegraphics[width=\textwidth]{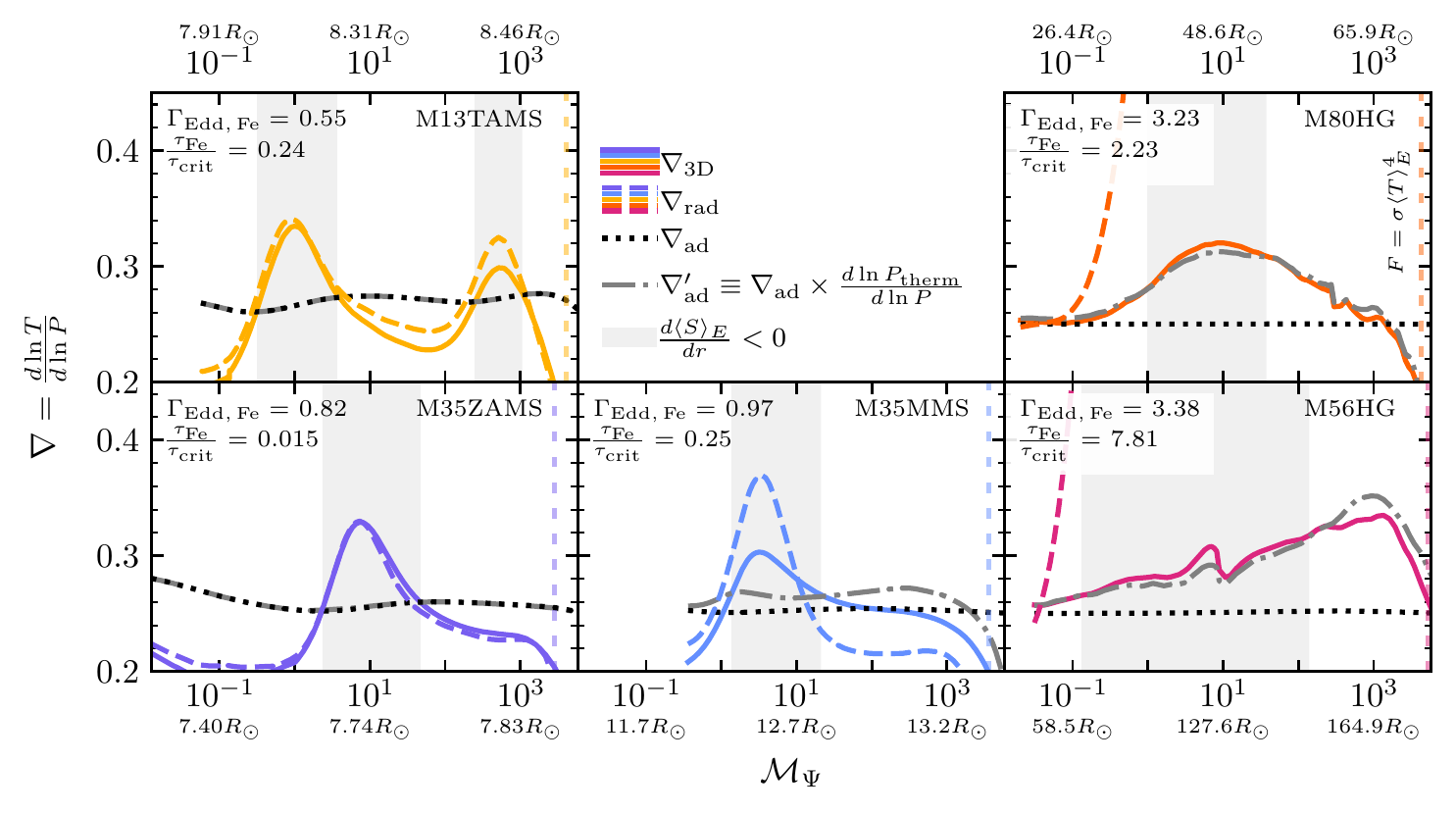} 
\vspace*{-0.75cm}
\caption{Comparison of temperature gradients for the five 3D RHD models including the realized temperature gradient, $\nabla_{\mathrm{3D}}$, radiative temperature gradient, $\nabla_{\mathrm{rad}}$, adiabatic gradient, $\nabla_{\mathrm{ad}}$, and the modified model adiabatic temperature gradient including $P_{\mathrm{turb}}$, $\nabla_{\mathrm{ad}}'$ (colored solid, colored dashed, black dotted, and grey dot-dashed lines respectively).
The vertical dashed lines are the same as in Figure~\ref{fig:mom_check}.
The grey shaded regions denote convectively unstable regions as in Figure~\ref{fig:new_Ts}.}
\label{fig:grads}
\end{center}
\end{figure*}

The MESA models utilizing the default MLT parameters also predict convective velocities at the Fe opacity peak.
Table~\ref{tab:3D_models} shows the difference between these velocities and the 3D RMS velocities at both the Fe opacity peak ($v_{\mathrm{MESA,Fe}}$ and $v_{\mathrm{Fe}}$ respectively) and the He opacity peak ($v_{\mathrm{MESA,He}}$ and $v_{\mathrm{He}}$ respectively).
The He opacity peak velocities disagree with those seen in the 3D models due to convective overshooting.
At the Fe opacity peak, the velocities differ by $32\%$ in M13TAMS, a factor of 3 in M35ZAMS, and $22\%$ in M35MMS.
Considering these comparisons utilize the same default MLT parameters for all the models, it is surprisingly accurate given the differences of the models. 

Modifying the $y$ and $\alpha$ MLT parameters for each MESA model can yield better agreement.
Keeping $\alpha=2$, the peak velocities in MESA's FeCZ agree when $y=0.2$, $y=0.1$, and $y=1$ for M13TAMS, M35ZAMS, and M35MMS respectively.
Alternatively, for $y=1/3$ we find agreement with $\alpha=1.7$, $\alpha=1.3$, and $\alpha=4.5$ for M13TAMS, M35ZAMS, and M35MMS respectively.
These changes in $y$ and $\alpha$ do not create significant changes in $T_{\mathrm{eff}}$, $r_{\mathrm{photo}}$, or the temperature profiles with all deviations being $<0.5\%$ as radiation remains the dominant energy transport mechanism.
Both $y$ and $\alpha$ impact the estimates of the convective velocity, however they impact different weak turbulence regimes.
Reducing $y$ is likely the correct modification when $\tau\ll\tau_{\mathrm{crit}}$ (M35ZAMS) while reducing $\alpha$ is the better choice in the lower luminosity regime where $L\ll L_{\mathrm{Edd}}$ (M13TAMS).

\subsection{Temperature Gradients in Turbulent Envelopes}
Energy transport defines the temperature gradient, $\nabla\equiv\ln{T}/d\ln{P}$, in 1D stellar models which is compared to two distinctly defined gradients.
The first is $\nabla_{\mathrm{rad}}$, the temperature gradient required to carry the flux solely by radiative diffusion \citep[see Eq.~27 of ][]{Henyey1965}.
The second is the adiabatic temperature gradient, $\nabla_{\mathrm{ad}}$, which is often realized in efficient convection as plumes rise and fall nearly adiabatically.
Figure~\ref{fig:grads} compares the 3D temperature gradient, $\nabla_{\mathrm{3D}}$, with $\nabla_{\mathrm{rad}}$ and $\nabla_{\mathrm{ad}}$.
For M13TAMS and M35ZAMS, $\nabla$ is well approximated by $\nabla_{\mathrm{rad}}$ further agreeing with the $\tau\ll\tau_{\mathrm{crit}}$ picture.
The convective plumes lose enough heat from radiation as they rise that the convective flux is negligible and thus $\nabla\approx\nabla_{\mathrm{rad}}$.

In more turbulent models, a different picture unfolds with $\nabla_{\mathrm{3D}}$ deviating from both $\nabla_{\mathrm{rad}}$ and $\nabla_{\mathrm{ad}}$.
Below the Fe opacity peak in M35MMS, $\nabla$ follows $\nabla_{\mathrm{rad}}$ to within $5\%$, however there is a departure from $\nabla_{\mathrm{rad}}$ in and above the CZ that is also seen in M80HG and M56HG.
Models M80HG and M56HG have $L\approx L_{\mathrm{Edd}}$ and exhibit $P_{\mathrm{r}}\gg P_{\mathrm{gas}}$ resulting in $\nabla_{\mathrm{rad}}\gg\nabla_{\mathrm{ad}}$.
As both are convectively efficient ($\tau\gtrsim\tau_{\mathrm{crit}}$) the convective flux grows until the velocities approach the sound speed, limiting the energy transport.
The resulting $\nabla_{\mathrm{3D}}$ lies between $\nabla_{\mathrm{rad}}$ and $\nabla_{\mathrm{ad}}$ suggesting a different mechanism of convective inefficiency, aside from radiative losses.

However as discussed earlier, these trans-sonic velocities imply a significant $P_{\mathrm{turb}}$, suggesting the need for a modified $\nabla_{\mathrm{ad}}$.
In this regime, \cite{Henyey1965} used the chain rule to account for the role of turbulent pressure in $\nabla_{\mathrm{ad}}$ by defining,
\begin{equation}
\nabla_{\mathrm{ad}}'=\nabla_{\mathrm{ad}}\times\frac{d\ln{P_{\mathrm{therm}}}}{d\ln{P}},
\end{equation}
where $P=P_{\mathrm{therm}}+P_{\mathrm{turb}}$.
For M13TAMS and M35ZAMS, $\nabla_{\mathrm{ad}}'$ is equivalent to $\nabla_{\mathrm{ad}}$ but for the others there is a noticeable change.
In fact, when $\nabla_{\mathrm{ad}}'$ is used, the superadiabaticity ($\nabla_{\mathrm{3D}}-\nabla_{\mathrm{ad}}'$) of M80HG and M56HG is reduced significantly implying the convection is truly efficient and follows the modified model adiabatic temperature gradient that includes the impact of $P_{\mathrm{turb}}$.
Additionally, the definition of the CZs, where $dS/dr<0$, agree identically with the regions where $\nabla>\nabla_{\mathrm{ad}}'$ rather than the conventional $\nabla_{\mathrm{ad}}$ in all the models.
Thus when models have significant turbulent pressure, 1D models must utilize $\nabla_{\mathrm{ad}}'$ to properly estimate the CZs' properties.

%%%%%%%%%%%%%%%%%%%%%%%%%%%%%%%%%%%%%%%%%%%%%%%%%%%%%%%%%%
\section{Conclusions}
We present five 3D RHD \Athena\ models that highlight the impact of turbulence in massive star envelopes.
When the stellar luminosities are sufficiently high and NSCZs are adequately optically thick as to develop trans-sonic convective velocities, the ram pressure of the turbulence can account for nearly $80\%$ of the pressure support.
This additional pressure support modifies the temperature change of an adiabitically rising plume, reducing the estimated superadiabaticity and confirming that convection is efficient.
Estimating the turbulent pressure for 1D models using physically motivated prescriptions will be vital for improving estimates of stellar observables.
That being said, it is important to remember that turbulence makes photospheres inherently 3D and care should be taken when interpreting 1D photospheric values in this realm. 

The presence of ubiquitous stochastic low frequency (SLF) brightness variability from massive stars has caused significant debate about their origin.
The SLF variability was shown to naturally originate from the NSCZs in the more massive simulations presented in this work (M35ZAMS, M35MMS, M80HG, M56HG) \citep{Schultz2022}.
However, the formation mechanism for lower mass and less luminous stars remains a mystery with core internal gravity waves \citep[e.g.][]{Bowman2020a} and stellar wind variability \cite[e.g.][]{Krticka2018} also proposed as possible sources.
Our introduction of the lower mass model M13TAMS allows us to address what occurs in lower mass stars.
Following \cite{Cantiello2021}, the characteristic frequency, $\nu_{\mathrm{char}}$, of M13TAMS is $1.5\,$d$^{-1}$ consistent with observations of similar stars \citep{Bowman2020a}.
The RMS amplitude of the $\log{L}$ fluctuations from M13TAMS (accounting for the $\sqrt{n}$ geometric factor of combining $n$ wedge models across the stellar surface) is $\approx130\,\mu$mag, comparable to the amplitude of variability in observed stars.
\hlm{Thus NSCZs in all five of our massive star envelope models exhibit SLF variability on timescales consistent with measured $\nu_{\mathrm{char}}$ and
at levels comparable to those observed.
The finite runtime of these 3D models limit our predictions of variability on timescales longer or comparable to the models' duration ($\lesssim 10\,$d).}

These 3D RHD models are computationally expensive (taking 3000 Skylake cores 4 days to run 1 model day), limiting our ability to populate the HRD and generate a $P_{\mathrm{turb}}$ prescription to be used in 1D stellar evolution. 
However our work has highlighted two clear regimes.
In the weak turbulence regime, in the lower left of the HRD, the NSCZs are either too radiatively lossy (M35ZAMS) or have sufficiently low luminosities (M13TAMS) to impact the stellar structure.
The opposite is true for the efficient luminous regime in the upper right of the HRD (M80HG and M56HG) where turbulence dominates the envelope.
M35MMS is the only model currently in the transition between these two regimes. 
For this reason we have only attempted to compare the existing models to their MESA counterparts and note the impacts of the turbulent NSCZs.
We are excited to see that 1D stellar evolution models agree with 3D RHD simulations in the weak turbulence regime and as more models are computed, the transition between spherically symmetric and inherently 3D envelopes may be mapped and 1D turbulent pressure prescriptions developed to improve stellar evolution models.

We thank Matteo Cantiello, Jared Goldberg, and Benny Tsang for many helpful conversations and comments.
This research was supported in part by the NASA ATP grant \hlm{ATP-80NSSC22K0725}, by the National Science Foundation through grant PHY 17-48958 at the KITP.
Resources supporting this work were also provided by the NASA High-End Computing (HEC) programme through the NASA Advanced Supercomputing (NAS) Division at Ames Research Center. 
We acknowledge support from the Center for Scientific Computing from the CNSI, MRL: an NSF MRSEC (DMR-1720256) and NSF CNS-1725797.
The Flatiron Institute is supported by the Simons Foundation.


\bibliography{sample631}{}

\begin{thebibliography}{}
\expandafter\ifx\csname natexlab\endcsname\relax\def\natexlab#1{#1}\fi
\providecommand{\url}[1]{\href{#1}{#1}}
\providecommand{\dodoi}[1]{doi:~\href{http://doi.org/#1}{\nolinkurl{#1}}}
\providecommand{\doeprint}[1]{\href{http://ascl.net/#1}{\nolinkurl{http://ascl.net/#1}}}
\providecommand{\doarXiv}[1]{\href{https://arxiv.org/abs/#1}{\nolinkurl{https://arxiv.org/abs/#1}}}

\bibitem[{{B{\"o}hm-Vitense}(1958)}]{BohmVitense1958}
{B{\"o}hm-Vitense}, E. 1958, \zap, 46, 108

\bibitem[{{Bowman} {et~al.}(2020){Bowman}, {Burssens}, {Sim{\'o}n-D{\'\i}az},
  {Edelmann}, {Rogers}, {Horst}, {R{\"o}pke}, \& {Aerts}}]{Bowman2020a}
{Bowman}, D.~M., {Burssens}, S., {Sim{\'o}n-D{\'\i}az}, S., {et~al.} 2020,
  \aap, 640, A36, \dodoi{10.1051/0004-6361/202038224}

\bibitem[{{Cantiello} {et~al.}(2021){Cantiello}, {Lecoanet}, {Jermyn}, \&
  {Grassitelli}}]{Cantiello2021}
{Cantiello}, M., {Lecoanet}, D., {Jermyn}, A.~S., \& {Grassitelli}, L. 2021,
  arXiv e-prints, arXiv:2102.05670.
\newblock \doarXiv{2102.05670}

\bibitem[{{Cantiello} {et~al.}(2009){Cantiello}, {Langer}, {Brott}, {de Koter},
  {Shore}, {Vink}, {Voegler}, {Lennon}, \& {Yoon}}]{Cantiello2009}
{Cantiello}, M., {Langer}, N., {Brott}, I., {et~al.} 2009, \aap, 499, 279,
  \dodoi{10.1051/0004-6361/200911643}

\bibitem[{{Couch} \& {Ott}(2015)}]{Couch2015}
{Couch}, S.~M., \& {Ott}, C.~D. 2015, \apj, 799, 5,
  \dodoi{10.1088/0004-637X/799/1/5}

\bibitem[{{Goldberg} {et~al.}(2021){Goldberg}, {Jiang}, \&
  {Bildsten}}]{Goldberg2021}
{Goldberg}, J.~A., {Jiang}, Y.-F., \& {Bildsten}, L. 2021, arXiv e-prints,
  arXiv:2110.03261.
\newblock \doarXiv{2110.03261}

\bibitem[{{Goldberg} {et~al.}(2022){Goldberg}, {Jiang}, \&
  {Bildsten}}]{Goldberg2022}
---. 2022, \apj, 929, 156, \dodoi{10.3847/1538-4357/ac5ab3}

\bibitem[{{Grassitelli} {et~al.}(2015){Grassitelli}, {Fossati},
  {Sim{\'o}n-Di{\'a}z}, {Langer}, {Castro}, \& {Sanyal}}]{Grassitelli2015}
{Grassitelli}, L., {Fossati}, L., {Sim{\'o}n-Di{\'a}z}, S., {et~al.} 2015,
  \apjl, 808, L31, \dodoi{10.1088/2041-8205/808/1/L31}

\bibitem[{{Henyey} {et~al.}(1965){Henyey}, {Vardya}, \&
  {Bodenheimer}}]{Henyey1965}
{Henyey}, L., {Vardya}, M.~S., \& {Bodenheimer}, P. 1965, \apj, 142, 841,
  \dodoi{10.1086/148357}

\bibitem[{{Jermyn} {et~al.}(2022){Jermyn}, {Anders}, {Lecoanet}, {Cantiello},
  \& {Goldberg}}]{Jermyn2022RN}
{Jermyn}, A.~S., {Anders}, E.~H., {Lecoanet}, D., {Cantiello}, M., \&
  {Goldberg}, J.~A. 2022, Research Notes of the American Astronomical Society,
  6, 29, \dodoi{10.3847/2515-5172/ac531e}

\bibitem[{{Jermyn} {et~al.}(2023){Jermyn}, {Bauer}, {Schwab}, {Farmer}, {Ball},
  {Bellinger}, {Dotter}, {Joyce}, {Marchant}, {Mombarg}, {Wolf}, {Sunny Wong},
  {Cinquegrana}, {Farrell}, {Smolec}, {Thoul}, {Cantiello}, {Herwig}, {Toloza},
  {Bildsten}, {Townsend}, \& {Timmes}}]{Jermyn2023}
{Jermyn}, A.~S., {Bauer}, E.~B., {Schwab}, J., {et~al.} 2023, \apjs, 265, 15,
  \dodoi{10.3847/1538-4365/acae8d}

\bibitem[{{Jiang}(2021)}]{Jiang2021}
{Jiang}, Y.-F. 2021, \apjs, 253, 49, \dodoi{10.3847/1538-4365/abe303}

\bibitem[{{Jiang} {et~al.}(2015){Jiang}, {Cantiello}, {Bildsten}, {Quataert},
  \& {Blaes}}]{Jiang2015}
{Jiang}, Y.-F., {Cantiello}, M., {Bildsten}, L., {Quataert}, E., \& {Blaes}, O.
  2015, \apj, 813, 74, \dodoi{10.1088/0004-637X/813/1/74}

\bibitem[{{Jiang} {et~al.}(2018){Jiang}, {Cantiello}, {Bildsten}, {Quataert},
  {Blaes}, \& {Stone}}]{Jiang2018}
{Jiang}, Y.-F., {Cantiello}, M., {Bildsten}, L., {et~al.} 2018, \nat, 561, 498,
  \dodoi{10.1038/s41586-018-0525-0}

\bibitem[{{Joss} {et~al.}(1973){Joss}, {Salpeter}, \& {Ostriker}}]{Joss1973}
{Joss}, P.~C., {Salpeter}, E.~E., \& {Ostriker}, J.~P. 1973, \apj, 181, 429,
  \dodoi{10.1086/152060}

\bibitem[{{Krti{\v{c}}ka} \& {Feldmeier}(2018)}]{Krticka2018}
{Krti{\v{c}}ka}, J., \& {Feldmeier}, A. 2018, \aap, 617, A121,
  \dodoi{10.1051/0004-6361/201731614}

\bibitem[{{Kuhfuss}(1986)}]{Kuhfuss1986}
{Kuhfuss}, R. 1986, \aap, 160, 116

\bibitem[{{Ludwig} {et~al.}(1999){Ludwig}, {Freytag}, \&
  {Steffen}}]{Ludwig1999}
{Ludwig}, H.-G., {Freytag}, B., \& {Steffen}, M. 1999, \aap, 346, 111.
\newblock \doarXiv{astro-ph/9811179}

\bibitem[{{Paxton} {et~al.}(2011){Paxton}, {Bildsten}, {Dotter}, {Herwig},
  {Lesaffre}, \& {Timmes}}]{Paxton2011}
{Paxton}, B., {Bildsten}, L., {Dotter}, A., {et~al.} 2011, \apjs, 192, 3,
  \dodoi{10.1088/0067-0049/192/1/3}

\bibitem[{{Paxton} {et~al.}(2013){Paxton}, {Cantiello}, {Arras}, {Bildsten},
  {Brown}, {Dotter}, {Mankovich}, {Montgomery}, {Stello}, {Timmes}, \&
  {Townsend}}]{Paxton2013}
{Paxton}, B., {Cantiello}, M., {Arras}, P., {et~al.} 2013, \apjs, 208, 4,
  \dodoi{10.1088/0067-0049/208/1/4}

\bibitem[{{Paxton} {et~al.}(2015){Paxton}, {Marchant}, {Schwab}, {Bauer},
  {Bildsten}, {Cantiello}, {Dessart}, {Farmer}, {Hu}, {Langer}, {Townsend},
  {Townsley}, \& {Timmes}}]{Paxton2015}
{Paxton}, B., {Marchant}, P., {Schwab}, J., {et~al.} 2015, \apjs, 220, 15,
  \dodoi{10.1088/0067-0049/220/1/15}

\bibitem[{{Paxton} {et~al.}(2018){Paxton}, {Schwab}, {Bauer}, {Bildsten},
  {Blinnikov}, {Duffell}, {Farmer}, {Goldberg}, {Marchant}, {Sorokina},
  {Thoul}, {Townsend}, \& {Timmes}}]{Paxton2018}
{Paxton}, B., {Schwab}, J., {Bauer}, E.~B., {et~al.} 2018, \apjs, 234, 34,
  \dodoi{10.3847/1538-4365/aaa5a8}

\bibitem[{{Paxton} {et~al.}(2019){Paxton}, {Smolec}, {Schwab}, {Gautschy},
  {Bildsten}, {Cantiello}, {Dotter}, {Farmer}, {Goldberg}, {Jermyn}, {Kanbur},
  {Marchant}, {Thoul}, {Townsend}, {Wolf}, {Zhang}, \& {Timmes}}]{Paxton2019}
{Paxton}, B., {Smolec}, R., {Schwab}, J., {et~al.} 2019, \apjs, 243, 10,
  \dodoi{10.3847/1538-4365/ab2241}

\bibitem[{{Ricker} {et~al.}(2015){Ricker}, {Winn}, {Vanderspek}, {Latham},
  {Bakos}, {Bean}, {Berta-Thompson}, {Brown}, {Buchhave}, {Butler}, {Butler},
  {Chaplin}, {Charbonneau}, {Christensen-Dalsgaard}, {Clampin}, {Deming},
  {Doty}, {De Lee}, {Dressing}, {Dunham}, {Endl}, {Fressin}, {Ge}, {Henning},
  {Holman}, {Howard}, {Ida}, {Jenkins}, {Jernigan}, {Johnson}, {Kaltenegger},
  {Kawai}, {Kjeldsen}, {Laughlin}, {Levine}, {Lin}, {Lissauer}, {MacQueen},
  {Marcy}, {McCullough}, {Morton}, {Narita}, {Paegert}, {Palle}, {Pepe},
  {Pepper}, {Quirrenbach}, {Rinehart}, {Sasselov}, {Sato}, {Seager},
  {Sozzetti}, {Stassun}, {Sullivan}, {Szentgyorgyi}, {Torres}, {Udry}, \&
  {Villasenor}}]{Ricker2015}
{Ricker}, G.~R., {Winn}, J.~N., {Vanderspek}, R., {et~al.} 2015, Journal of
  Astronomical Telescopes, Instruments, and Systems, 1, 014003,
  \dodoi{10.1117/1.JATIS.1.1.014003}

\bibitem[{{Schultz} {et~al.}(2020){Schultz}, {Bildsten}, \&
  {Jiang}}]{Schultz2020}
{Schultz}, W.~C., {Bildsten}, L., \& {Jiang}, Y.-F. 2020, \apj, 902, 67,
  \dodoi{10.3847/1538-4357/abb405}

\bibitem[{{Schultz} {et~al.}(2022){Schultz}, {Bildsten}, \&
  {Jiang}}]{Schultz2022}
---. 2022, \apjl, 924, L11, \dodoi{10.3847/2041-8213/ac441f}

\bibitem[{{Schultz} {et~al.}(2023){Schultz}, {Tsang}, {Bildsten}, \&
  {Jiang}}]{Schultz2023}
{Schultz}, W.~C., {Tsang}, B. T.~H., {Bildsten}, L., \& {Jiang}, Y.-F. 2023,
  \apj, 945, 58, \dodoi{10.3847/1538-4357/acb701}

\bibitem[{{Smith} {et~al.}(2004){Smith}, {Vink}, \& {de Koter}}]{Smith2004}
{Smith}, N., {Vink}, J.~S., \& {de Koter}, A. 2004, \apj, 615, 475,
  \dodoi{10.1086/424030}

\bibitem[{{Stone} {et~al.}(2020){Stone}, {Tomida}, {White}, \&
  {Felker}}]{Stone2020}
{Stone}, J.~M., {Tomida}, K., {White}, C.~J., \& {Felker}, K.~G. 2020, \apjs,
  249, 4, \dodoi{10.3847/1538-4365/ab929b}

\end{thebibliography}
\bibliographystyle{aasjournal}



\end{CJK*}
\end{document}